\documentstyle[12pt]{article}
\begin{document}
\setlength{\baselineskip}{0.30in}
\newcommand{\be}{\begin{eqnarray}}
\newcommand{\ee}{\end{eqnarray}}
\newcommand{\bi}{\bibitem}
\newcommand{\nt}{\nu_\tau}
\newcommand{\nm}{\nu_\mu}
\newcommand{\nue}{\nu_e}
\newcommand{\mnt}{m_{\nu_\tau}}
\newcommand{\mnm}{m_{\nu_\mu}}
\newcommand{\mne}{m_{\nu_e}}

{\begin{center}
\vglue .06in
{\Large \bf { Nonequilibrium cosmic neutrinos and nucleosynthesis.
  }
}
\bigskip
\\{\bf A.D. Dolgov}
 \\[.05in]
{\it{Teoretisk Astrofysik Center\\
 Juliane Maries Vej 30, DK-2100, Copenhagen, Denmark
\footnote{Also: ITEP, Bol. Cheremushkinskaya 25, Moscow 113259, Russia.}
}}

\end{center}
\begin{abstract}

The neutrino role in primordial nucleosynthesis is reviewed. The importance of
nonequilibrium effects is emphasized both for the standard massless and
possibly massive neutrinos. The upper bound on tau neutrino mass is presented.
A spatial variation of primordial abundances and a possibility of observing
them by precise measurements of the CMB anisotropy are considered.The
nucleosynthesis bounds on the parameters of neutrino oscillations into
sterile neutrinos are discussed.

\end{abstract}

\section{Introduction}

It is well known that neutrinos have a very important impact on
cosmology (for a recent review and a list of references see e.g. \cite{ad}).
In particular, the comparison of primordial nucleosynthesis theory with
observational data permits to put rather stringent bounds on neutrino
properties, their mass, number of flavors, possible new interactions, etc.
To this end a detailed study of neutrino kinetics
in the primeval plasma, especially nonequilibrium corrections, which happen to
be quite essential, is of primary importance and the proper treatment of the
latter significantly changes the results of the simpler equilibrium
calculations. Technically the problem is very complicated and demands an
accurate numerical solution of a system of coupled integro-differential
kinetic equations, however in many cases a rough order-of-magnitude
estimate of non-equilibrium corrections can be done analytically.

Surprisingly nonequilibrium corrections to the energy spectrum are not very
small even for massless neutrinos in the standard model and of course they
are very large in the case of possibly heavy $\nt$ with the mass in MeV range.
Account of nonequilibrium effects permits to considerably improve the
nucleosynthesis bounds on the mass of $\nt$.

Another important effect, where deviations from the standard equilibrium
non-degenerate Fermi-Dirac distribution
\be
f(p,t)= {1 \over \exp[E /T(t)] + 1 }
\label{ffd}
\ee
is essential for nucleosynthesis, is a possible lepton asymmetry. The latter
could be either primordial, generated at a very early stage, or it may arise
during nucleosynthesis epoch due to nonequilibrium neutrino oscillations. Even
if lepton asymmetry is not generated by oscillations, they still might have
a very important impact on nucleosynthesis and the study of nucleosynthesis
leads to interesting bounds on oscillation parameters.

If there are some new particles, abundant in the plasma during nucleosynthesis,
they would change the cooling rate of the plasma and thus change the
standard abundances of the light elements. Normally the effect of such new
particles is just to change the expansion/cooling rates but in some cases
their interaction may also produce nonequilibrium neutrinos, especially $\nue$,
and in this case the impact on nucleosynthesis would be significantly
different.

In what follows I briefly review these subjects. In section 2 nonequilibrium
corrections to the spectra of normal massless neutrinos and their possible
observational manifestations are discussed. In section 3 the role of a possibly
heavy $\nt$ in nucleosynthesis is considered and an upper bound on its mass
is presented. Neutrino degeneracy and especially a possible variation of the
latter on the cosmological scales (a few 100 Mpc or even Gpc) is discussed in
section 4. In section 5 neutrino oscillations are considered. In Conclusion
the main results of this brief review is summarized.

\section{Nonequilibrium massless neutrinos}

It is  usually assumed that thermal relics with $m=0$ are in perfect
equilibrium state even after decoupling. For the photons in cosmic microwave
background (CMB) it is known with a very high accuracy. The same assumption
is made about neutrinos so that their distribution is given by
eq.~(\ref{ffd}). Indeed when the interaction rate is high in comparison with
the expansion rate, $\Gamma_{int} \gg H$, the equilibrium is evidently
established. When interactions can be neglected the distribution function
may have an arbitrary form but for massless particles the equilibrium
distribution is preserved if it was established earlier at a dense and hot
stage when the interaction was fast. One can see that from kinetic equation
in the expanding universe:
\be
(\partial_t - Hp\partial_p) f_j (p_j,\,t) = I^{coll}_{j}
\label{dtfj}
\ee
where the collision integral in the r.h.s. vanishes for the equilibrium
functions:
\be
f^{(eq)} = \left( e^{E/T - \mu/T} \pm 1\right)^{-1}
\label{feq}
\ee
The temperature $T$ and  chemical potential $\mu$ may be functions
of time.

The l.h.s. is annihilated by $f = f^{(eq)}$ if the following condition is
fulfilled for any value of particle energy $E$ and
momentum $p=\sqrt{E^2 -m^2}$:
\be
{\dot T \over T} + H {p\over E} {\partial E \over \partial p}
- {\mu \over E}\left( {\dot \mu \over \mu} -{\dot T \over T}\right) =0
\label{dott}
\ee
This can only be true if $p=E$ (i.e $m=0$), $\dot T/T = -H$, and
$\mu \sim T$. It can be shown that for massless particles, which initially
possessed equilibrium distribution, temperature and
chemical potential indeed satisfy these requirements for $I^{coll} =0$,
so the equilibrium distribution is not destroyed even when the interaction
is switched off.

It  would be true for neutrinos if they instantly decoupled from the
electromagnetic component of the plasma (electrons, positrons, and photons)
at the moment when neutrino interactions was strong enough so that at the
moment of decoupling they were in thermally equilibrium state with the
same temperature as photons and $e^{\pm}$. According to simple estimates
the decoupling temperature, $T_{dec}$, for $\nue$ is about 2 MeV and that
for $\nm$ and $\nt$ is about 3 MeV.
In reality the decoupling is not instantaneous and even below $T_{dec}$ there
is some residual interaction between $e^{\pm}$ and neutrinos. An important
point is that after neutrino decoupling the temperature of the electromagnetic
component of the plasma became somewhat higher than the neutrino temperature.
The electromagnetic part of the plasma is heated by the annihilation of
{\it massive} electrons and positrons. This is the well known effect
which ultimately results in the present day ratio of temperatures,
$T_\gamma /T_\nu = (11/4)^{1/3}$.
During primordial nucleosynthesis the temperature difference between
electromagnetic and neutrino components of the plasma is small but still
non-vanishing. Due to this temperature difference the annihilation of the
hotter electrons/positrons, $e^+ e^- \rightarrow \bar \nu \nu$, heats up
the neutrino component of the plasma and distorts neutrino spectrum. The
average neutrino heating under assumption of their equilibrium spectrum
was estimated in refs. \cite{hh,rm}. Spectrum distortion in Boltzmann
approximation was calculated numerically in ref. \cite{dt} and analytically
in \cite{df}. In accordance with the latter it takes the form:
\be
{\delta f_{\nue} \over f_{\nue} }\approx 3\cdot 10^{-4} \,\,{E\over T}
\left( {11 E \over 4T } - 3\right)
\label{dff}
\ee
The distortion of the spectra of $\nm$ and $\nt$ is approximately twice weaker.
Here $\delta f = f - f^{(eq)}$.

An exact numerical treatment of the problem was first done in ref. \cite{hm}
and later, with a better precision and a corrected expression for the matrix
element of one of the participating reactions, in ref. \cite{dhs0}.
The system of coupled kinetic equations (\ref{dtfj}) governing the neutrino
distribution functions with the collision integral of the form
\be
I_{coll} = {1\over 2E_1}\sum \int {d^3 p_2 \over 2E_2 (2\pi)^3}
{d^3 p_3 \over 2E_3 (2\pi)^3}{d^3 p_4 \over 2E_4 (2\pi)^3}
\nonumber \\
(2\pi)^4\delta^{(4)} (p_1+p_2-p_3-p_4) F(f_1,f_2,f_3,f_4)
S\, |A|^2_{12\rightarrow 34}
\label{icoll}
\ee
was solved numerically with the precision about $10^{-4}$. In the
expression (\ref{icoll})
\be
F(f_1,f_2,f_3,f_4) = f_3 f_4 (1-f_1)(1-f_2)-f_1 f_2 (1-f_3)(1-f_4)
\label{F}
\ee
and $ |A|^2_{12\rightarrow 34}$ is the matrix element squared of the
4-fermion weak interaction. The results of ref. \cite{dhs0} confirmed
the shape of spectrum distortion (\ref{dff} ). The total relative change in
neutrino energy densities was found to be
\be
\delta \rho_{\nue} / \rho_{0} = 0.9 \%, \,\,\,
\delta \rho_{\nm,\nt} / \rho_{0} = 0.4 \%,
\label{drho}
\ee
where $\rho_{0}$ is the unperturbed neutrino energy density.

Naively one would expect that distortion of neutrino energy density at a
per cent level would result in the similar distortion in the primordial
abundances of light elements. However this does not occur by the following
reason. An excess of neutrinos at high energy tail of the spectrum results
in excessive destruction of neutrons in the reaction:
\be
n+\nu_e \leftrightarrow p + e^-
\label{nnue}
\ee
and an excessive creation in the reaction:
\be
n+e^+ \leftrightarrow p +  \bar \nu
\label{ne}
\ee
This nonequilibrium contribution into the second process is more efficient
because the number density of protons at nucleosynthesis (when $T\approx 0.7$
MeV) is 6-7 times larger than that of neutrons. So
an excess of high energy neutrinos results in an increase of
the frozen neutron-to proton ratio, $r$, and in the
corresponding increase of $^4 He$. On the other hand an excess of
of neutrinos at low energies results in a decrease of $r$ because reaction
(\ref{ne}) is suppressed due to threshold effects. It happened that
the discussed above nonequilibrium spectrum distortion took place in the
middle between the two extremes and the net influence of these distortion
on e.g. $^4 He$ is quite small, the change of the mass fraction of $^4 He$ is
$\sim 10^{-4}$.

Though quite small, such extra heating of neutrinos may be in principle
noticed in future high precision measurements of CMB
anisotropies~\cite{ldht,gg}. A change in neutrino energy density with respect
to the standard case would result in a shift of equilibrium epoch between
matter and radiation, which is imprinted on the form of the angular spectrum
of fluctuations of CMB. Because of potential observability of
distortion of neutrino energy density it
was recalculated in ref. \cite{gg} where a larger result, than found
in the previous papers, was obtained. In this connection we \cite{dhsa}
repeated our calculations with a larger number of integration points,
and wider momentum range, checked the
stability of our calculation procedure and confirmed our
previous results \cite{dhs0} with the precision of about $10^{-4}$. One
possible source of disagreement may be an incorrect probability of the
reactions $\nu_a+\nu_a\rightarrow \nu_a+\nu_a$ ($a=e,\mu,\tau$) used in
ref. \cite{gg} and a smaller number of integration points in the essential
region.

\section{Nonequilibrium massive $\nt$ and nucleosynthesis}

It is well known that comparison of calculated primordial abundances of
light elements with observations permits to put a constraint
on the expansion/cooling rate at the primordial nucleosynthesis (NS) epoch.
In particular such arguments allow to limit the number of possible neutrino
species (or other particles abundant at NS) \cite{ht,pee,sch,ssg}. The
present day data seem to exclude one extra neutrino species and possibly even
0.3 (for a recent review and analysis see e.g. ref. \cite{gst}).

Similar arguments permit to put a stringent bound on the mass of $\nt$,
considerably better than the existing direct experimental limit,
$\mnt < 18 $ MeV \cite{mnutau}. Using this result and nucleosynthesis data
one can conclude that $\mnt < 0.5-1$ MeV, if such neutrino is stable at
the nucleosynthesis time scale, i.e. $\tau_{\nu_\tau} > 100 $ sec.

Equilibrium energy density of massive particles is smaller than the energy
density of massless ones. So if it was the case, then massive $\nt$ would
effectively correspond to a smaller number of massless neutrinos. However
when expansion rate $H = \dot a /a$ becomes smaller than the rate of
$\nt$-annihilation, equilibrium is no more maintained and the number and
energy densities of $\nt$ become larger than their equilibrium values. Because
of that one massive $\nt$ could correspond to several massless neutrino
species. Original calculations of the bound on $\mnt$ \cite{ktcs} were
made under the simplifying assumptions of Boltzmann statistics and kinetic
equilibrium of all participating particles. In other words the distribution
of massless $\nue$ and $\nt$ were taken as $f = \exp (-E/T)$, while the
distribution of massive $\nt$ were assumed to have the form:
\be
f_{\nt} = \exp (-E/T + \xi)
\label{fnt}
\ee
where the dimensionless
(pseudo)chemical potential $\xi$ is a function of time only and
does not depend on the particle momentum. (If lepton asymmetry is vanishingly,
small the values of $\xi$ for particles and antiparticles are the same.) In
this approximation the complicated integro-differential
equations~ (\ref{dtfj},\ref{icoll}) are reduced to
the well known ordinary differential equation \cite{zop}:
\be
\dot n_\nu +3H n_\nu = \langle \sigma_{ann} v \rangle
(n^{(eq)2}_{\nu} -n^2_\nu)
\label{dotn}
\ee
Here $n^{(eq)}$ is the equilibrium number density, $v$ is the velocity of
annihilating particles, and angular brackets mean thermal averaging.

The assumption of kinetic equilibrium (\ref{fnt}) is fulfilled if
the rate of elastic scattering at the moment of annihilation freezing,
$\Gamma_{ann} \sim H$, is much higher than both the expansion rate, $H$, and
the rate of annihilation, $\Gamma_{ann}$. It is generally correct because
the cross-sections of annihilation and elastic scattering are usually of
similar magnitudes but the rate of annihilation,
$\Gamma_{ann} \sim \sigma_{ann} n_m$ is suppressed relative to the rate
of elastic scattering, $\Gamma_{el} \sim \sigma_{el} n_0$, due to Boltzmann
suppression of the number density of massive particles, $n_m$, with respect
to that of massless ones, $n_0$. However in the case of MeV-neutrinos both
rates $\Gamma_{ann}$ and $\Gamma_{el}$
at the moment of annihilation freezing are of the same order of
magnitude. Correspondingly assumption of kinetic equilibrium at annihilation
freezing is strongly violated. A semi-analytic calculations of the
deviations from kinetic equilibrium were done in ref. \cite{ad1}, where
a perturbative approach was developed. In the case of a momentum-independent
amplitude of elastic scattering the integro-differential kinetic equation
in the Boltzmann limit can be reduced to the following differential equation:
\be
 JC'' + 2 J'C' =
- { 64\pi^3 H x^2 \over |A_0|^2 m} e^{y/2} \partial_y \left\{ e^{-y}
\partial_y\left[ e^{(u+y)/2} uy\partial_x \left(Ce^{-u}\right)\right]\right\}
\label{jc''}
\ee
where  $x=m/T$, $y=p/T$, prime means differentiation with respect to $y$,
$C(x,y)= \exp(\sqrt{x^2+y^2})f_m(x,y)$, and $f_m$ is the unknown
distribution function of massive particles.

A direct application of perturbation
theory (with respect to small deviation from equilibrium) to the
integro-differential kinetic equation (\ref{dtfj}) is impossible or very
difficult because the momentum dependence of the anzats for the first
order approximation to $f(p,t)$ is not known. Numerical solution of
exact kinetic equations \cite{dhsm} shows a reasonable agreement with
the semi-analytic approach based on eq. (\ref{jc''}).

It can be easily shown that the spectrum of massive $\nt$ is softer (colder)
than the equilibrium one. Indeed if elastic scattering of $\nt$, which would
maintain kinetic equilibrium is switched-off, the nonrelativistic $\nt$
cool down as $1/a^2$, while relativistic particles cool as $1/a$, where
$a(t)$ is the cosmological scale factor. Since the cross-section of
annihilation by the weak interactions is proportional
to the energy squared of the
annihilating particles, the annihilation of nonequilibrium $\nt$ is less
efficient and their number density becomes larger than in the equilibrium case.

After the pioneering paper \cite{ktcs} the frozen number density of $\nt$
was calculated with increased accuracies in refs. \cite{dr,hm1,dhsm}. The
better was the accuracy the larger was the
calculated frozen number density $n_{\nt}^{(f)}$.
In the maximum of $n_{\nt}^{(f)}$, which occurs
at $\mnt \approx 5$ MeV, the difference between the calculations of
the papers \cite{ktcs} and \cite{dhsm} is almost 50\%. So the account of
nonequilibrium effects in the distribution of massive $\nt$ results in
a larger frozen number density of $\nt$ and in a stronger influence on
primordial nucleosynthesis.

Another nonequilibrium effect is an extra cooling of massless $\nue$ and
$\nm$ due to their elastic scattering on colder $\nt$,
$\nu_{e,\mu}+\nt \rightarrow  \nu_{e,\mu}+\nt$. Because of that the inverse
annihilation $\nu_{e,\mu}+\bar \nu_{e,\mu}\rightarrow\nt+\nt$ is weaker
and the frozen number density of $\nt$ is smaller. But this is a second
order effect and is relatively unimportant.

Considerably more important is an overall
heating and modification of the spectrum of $\nue$
(and of course of $\bar \nue$) by the late annihilation
$\nt+\nt\rightarrow \nue +\bar \nue$ (the same is true for $\nm$ but
electronic neutrinos are more important for nucleosynthesis because they
directly participate in the reactions (\ref{nnue},\ref{ne}) governing the
frozen $n/p$-ratio. It is analogous to the similar effect originating from
$e^-e^+$-annihilation, considered in the previous section, but significantly
more profound. The overfall heating and spectral distortion work in the
opposite way for $\mnt> 1$ MeV. An overall increase of the number and
energy densities of $\nue$ and $\bar \nue$ result in a smaller temperature of
neutron freezing and in a decrease of the $n/p$-ratio. On the other hand
a hotter spectrum of $\nue$ shifts this ratio to a large value, as discussed
in the previous section. The latter effect was estimated semi-analytically in
ref. \cite{dpv}, where it was found that e.g. for $\mnt=20$ MeV the spectral
distortion is equivalent to 0.8 extra neutrino flavors for Dirac $\nt$ and to
0.1  extra neutrino flavors for Majorana $\nt$. The effect of overall heating
was found to be somewhat more significant \cite{fko,dhsm}.

Though the frozen number density of $\nt$ obtained in ref. \cite{dhsm}
is the largest (in comparison to the results of refs. \cite{ktcs,dr,hm1}),
the influence of nonequilibrium corrections on nucleosynthesis found
in \cite{dhsm} is somewhat weaker than that found in \cite{fko,hm1}
in the mass range above 15 MeV. It is
possibly related to a larger momentum cut-off in numerical calculations of
ref.~\cite{dhsm}, which gives rise to a smaller neutron freezing temperature.
For the graphical presentation of the results and comparison
with the other papers one can address ref. \cite{dhsm}.
The results of all nonequilibrium calculations are systematically and
considerably larger than those of the equilibrium ones of ref. \cite{ktcs}.
These newer and more accurate calculations permit to close the window
in the mass range 10-20 MeV, which is not excluded by nucleosynthesis if
the permitted number of extra neutrinos flavors is 1. Now even if 1 extra
neutrino is permitted, the upper bound on $\mnt$ is about 1 MeV. If 0.3
extra neutrino flavors are allowed, the $\nt$ mass is bounded from above by
0.3 MeV. These results are valid for the Majorana $\nt$. For the Dirac case
the mass bound from SN1987 is much more restrictive and, moreover the
calculations of nucleosynthesis limit on the mass of Dirac $\nt$ are
considerably more complicated because of a larger number of independent
unknown distribution functions.

\section{Lepton asymmetry and possible spatial variation of primordial
abundances.}

In the standard nucleosynthesis calculations is usually assumed that
neutrinos are not degenerate, or in other words, that their chemical
potentials are vanishing and their distributions are given by the
expression (\ref{ffd}). A justification for this assumption is a
small value of the baryon asymmetry, but strictly speaking very little
is known neither from observation nor theoretically about lepton asymmetry.
The best observational bounds are
found from primordial nucleosynthesis (for a recent reference
see e.g. \cite{kks}). Theoretically lepton asymmetry could be as small
as the baryon one, especially in the models with $(B-L)$-conservation, but
it also may be as large as unity \cite{dk,ftv,ccg}. Moreover, the asymmetry
could be not only large but also varying by unity at astronomically large
scales \cite{dk}.

The recent data \cite{tfb,rh,tbk,swc,dt1,wcl}
though rather controversial, may possibly indicate that the abundance of
primordial deuterium changes at the scales of the order of a gygaparsec or
several hundred megaparsecs. If the effect is real, there could be two
possible explanations of it. First, baryon asymmetry of the universe may be
not a universal constant but a varying function of space points \cite{jf,cos}.
This possibility meets certain problems with the primordial
$^7 Li$-abundance~\cite{jf} or with the isotropy of CMB \cite{cos}. Below
we discuss another possible source of a possible variation of primordial
abundances, namely spatially varying lepton asymmetries \cite{dp}.

It is noteworthy that independently of the data and theory, there is a
question what is known about light element abundances at large distances.
For example what is the upper or lower limit on $R_p$, the mass fraction of
primordial $^4 He$, at the distances above 100 Mpc? Is $R=50-60\%$
or even close to 100\% excluded? What is the characteristic scale where a
large variation of primordial abundances are permitted?
It is known from observations that the universe is (or better to say
was at the early stage) very homogeneous energetically. From isotropy
of CMB it follows that
\be
{\delta \rho \over \rho} < ({\rm a \,\,\,few}) \times 10^{-5}
\label{drho1}
\ee
A natural implication of the energetical homogeneity is the chemical
homogeneity but it is not necessarily so. It is interesting to consider a
model which gives rise to a small cosmological energy variation but to a
large chemical variation.

We assume that chemical potential of neutrinos, especially of $\nue$,
are varying on the scales above a few hundred Mpc. To explain the possibly
observed variation of deuterium, the dimensionless chemical potential of
electronic neutrinos, $\xi_{\nue}$ should vary by approximately unity. For
example $\xi_{\nue} =0$ in our neighborhood and $\xi_{\nue} = -1$ in
deuterium rich regions. With such variation of electronic asymmetry we
immediately obtain ${\delta \rho  / \rho} \approx
({\rm a\,\,\,few}) \times 10^{-3}$, much larger than the bound (\ref{drho}). To
save the model one has to assume a kind of lepton conspiracy \cite{dp}, namely
if in some space region of the universe lepton asymmetry is given by the
set of chemical potentials:
\be
\left\{ \xi_{\nue},\xi_{\nm},\xi_{\nt} \right\}
= \left\{ \alpha,\beta,\gamma\right\},
\label{abg}
\ee
then in another space region the asymmetry is given by a permutation of
$\alpha$, $\beta$, and $\gamma$. In this case the variation of the cosmological
energy density would vanish in the first approximation. Though the assumption
of lepton conspiracy looks as an a quite strong and artificial fine-tuning,
it can be rather naturally realized due the flavor symmetry,
$e \leftrightarrow \mu \leftrightarrow  \tau$.

It can be easily checked that if the variation of $^2 H$ is created by the
variation of $\xi_{\nue}$ from 0 to (-1), the corresponding mass fraction
of $^4 He$ in deuterium rich regions should be larger than 50\%~\cite{dp}.
Such a large variation of helium mass fraction would result to a considerable
density fluctuations due to different binding energies of helium and
hydrogen. Rescaling the estimates of ref. \cite{cos} one can find \cite{dp}
for the fluctuations of the CMB temperature :
\be
{\delta T \over T }\approx 10^{-5} \left({R_{hor} \over 10\lambda } \right)
\label{dtt}
\ee
where $\lambda$ is the wavelength of the fluctuation and $R_{hor}$ is the
present day horizon size. The restriction on the amplitude of temperature
fluctuations would be satisfied if $\lambda > 200 - 300 {\rm Mpc}/h_{100}$
($h_{100} = H/100$ km/sec/Mpc). Surprisingly direct astrophysical effects of
such big fluctuations of $R_p$ at distances above 100 Mpc cannot be
observed presently, at least the evident simple ones.

Another possibly dangerous effect is the differential neutrino heating
considered in section 2. If chemical potentials of neutrinos are different in
different space points, their nonequilibrium heating by $e^+e^-$-annihilation
would also be different. Correspondingly the photon temperature would also
be different. This effect was estimated in ref. \cite{dp}, where it was
found that ${\delta T / T }\approx 2 \times 10^{-5}$ for
$\delta \xi_\nu =1$.

A variation of mass fraction of primordial $^4 He$ could be observed in the
future high precision measurements of CMB anisotropies at small angular
scales \cite{hssw}. There are two possible effects, first, a slight
difference in recombination temperature which logarithmically depends on
hydrogen-to-photon ratio, and second, a strong suppression of high
multipoles with an increase of $R_p$. The latter is related to the
earlier helium recombination with respect to hydrogen and correspondingly to
a smaller number of free electrons at the moment of hydrogen recombination.
This in turn results in an increase of the mean free path of photons in
the primeval plasma and in a stronger Silk damping. The position and
the magnitude of the first acoustic peak remains practically
unchanged~\cite{hssw}.

This effect seems to be very promising for obtaining a bound on or an
observation of a possible variation of primordial helium mass fraction. If
this is the case then the amplitude of high multipoles at
different directions on the sky would be quite different.
The impact of the possible variation of primordial abundances on the
angular spectrum of CMB anisotropy at low $l$ is more model dependent. It
may have a peak corresponding to the characteristic scale
$R > 200-300$ Mpc or a plateau, which would mimic the effect of the
hot dark matter.

\section{Neutrino oscillations and nucleosynthesis}

An influence of oscillating neutrinos on nucleosynthesis depends
on possible oscillation channels. If the oscillations do not create any new
neutrino states, and if the initial (generated in the early universe)
lepton asymmetry is small, the impact of the oscillations on nucleosynthesis
is negligible. In the case
of nonzero lepton asymmetry the oscillations between $\nue$, $\nm$ and
$\nt$ (and their antiparticles) would result in a mixing of the different
lepton numbers. So that if the oscillations were fast enough at
nucleosynthesis (NS) and the equilibrium was established, all chemical
potentials would be equal. If the oscillations at NS were slow, then the
asymmetries would not be equalized and due to different refraction indices
for particles and antiparticles (see below) there might be even a
significant amplification of asymmetries.

A more interesting for NS effect takes place if neutrino oscillations
produce new neutrino states, e.g. a sterile neutrino (or neutrinos).
It may happen if the neutrino mass matrix contains both Dirac and Majorana
mass terms \cite{ad2}. In that case the sterile neutrino(s) is (are) just
the usual neutrino(s) with a wrong (positive) helicity induced by the Dirac
mass. If the characteristic time of oscillations is
sufficiently small, so that thermal equilibrium with respect to formation
of new states is fulfilled, there would be one or several new neutrino
species in the plasma and the only effect on NS is the corresponding change
in the expansion rate. It is mentioned above that one additional neutrino
species is forbidden by NS. This condition permits to exclude a certain range
of the oscillation parameters. In the original treatment of ref. \cite{ad2}
the influence of the medium on neutrino oscillations was neglected. In this
case the characteristic time of oscillations is just the vacuum time:
\be
\tau_{osc} = {E \over \delta m^2} =
10^{-3} \,\,{\rm sec}\,\,{ E/ {\rm MeV} \over
\delta m^2 /10^{-6}\,\,{\rm eV}^2 }
\label{tosc}
\ee
The rate of production of new neutrino species is
$\Gamma_{osc} = \left(\tau_{osc} \sin^2 2\theta\right)^{-1}$.
If $\Gamma_{osc} \geq H$ then the extra neutrino species would be abundantly
produced.

However for a large and interesting interval of masses and mixing angles
the influence of the medium cannot be neglected and one should take into
account
that neutrino refraction index in the primeval plasma at NS epoch is not
unity \cite{nr}:
\be
n^{\pm} -1 = \pm C_1 \eta_L {G_F T^3 \over E} + C_2 {G^2_F T^4 \over \alpha}
\label{n1}
\ee
where numerical coefficients $C_j$ are of order unity, $G_F$ is the Fermi
coupling constant, $\alpha= 1/137$ is the fine structure constant,
$E$ is the neutrino energy, $T$ is their temperature,
and $\eta_L$ is the leptonic asymmetry of the plasma. There can be different
asymmetries for different leptonic charges, then the expression above
should be correspondingly changed.

Neutrino oscillations with the account of dispersion effect were considered
in refs. \cite{bd1,bd2,kk1,ekm1,ekt,ssf}. It was shown that the oscillation
parameters are roughly speaking bounded by
\be
\sin^4 \theta |\delta m^2| < 10^{-2} {\rm eV}^2, \,\,\,
{\rm if} \sin^2 \theta < 0.1
\label{sin4}
\ee
and
\be
|\delta m|^2 < 10^{-6} {\rm eV}^2, \,\,\,{\rm if} \sin^2 \theta \approx 1
\label{dm2}
\ee
More recent calculations \cite{shi,chk,chk1} led to further clarification of
the bounds. It was shown in particular \cite{chk} that spectral distortion of
oscillating neutrinos, neglected in earlier calculations, is quite essential
for accurate determination of the changes in primordial abundances due to
oscillations.

A very interesting effect may take place if the MSW-resonance condition is
fulfilled for oscillations of neutrinos into sterile species. From the
expression for the refraction index (\ref{n1}) one can see that the resonance
condition is fulfilled either for neutrinos or anti-neutrinos depending on the
sign of the mass difference. If for example the transition of neutrinos into
sterile component is enhanced, then the leptonic asymmetry in the sector of
the usual (not sterile) neutrinos would rise up
and the oscillation would become more efficient, in turn producing more
asymmetry. The equation for asymmetry generation has the form
\be
\dot L = +AL
\label{dotl}
\ee
where $L$ is the lepton asymmetry and $A$ is a positive coefficient. When the
back reaction of the oscillation on the initial state of the plasma can be
neglected, the asymmetry rises up exponentially and can reach the values close
to unity. This effect was noticed in refs. \cite{bd1,ekm1} and the detailed
calculations showing that the effect can be quite large was done in
ref. \cite{ftv}. If e.g. the asymmetry is generated in electronic charge, the
impact on primordial nucleosynthesis would be quite significant and in
particular the limits on oscillation parameters might be less restrictive.
Still the mixing angles close to one and relatively large mass differences,
$\delta m^2 > 10^{-3}$, are forbidden if less than one extra neutrino
species is allowed by NS.

\section{Conclusion}

We see that nonequilibrium neutrino kinetics is quite essential at
nucleosynthesis. Even for the usual massless neutrinos the deviations from
equilibrium are rather large, at a per cent level. Though it has very little
effect on primordial abundances of light elements, about $10^{-4}$, the
corresponding changes in neutrino energy density may be in principle observed
in the future high precision measurement of angular variation of cosmic
microwave background.

Much more significant are nonequilibrium corrections to the spectra of possibly
massive tau-neutrinos if their mass lays in MeV region. Exact calculations
with all nonequilibrium corrections differ from the simpler equilibrium ones
as much as by 50\%. The nonequilibrium results are more restrictive and permit
to close a window for $\nt$ mass near 15 MeV which existed in equilibrium
calculations if primordial nucleosynthesis allowed for one extra neutrino
species. This permits to put the upper bound on Majorana type  mass of $\nt$
down to approximately 1 MeV.

Despite the fact that the early universe was very smooth energetically,
it is not excluded that chemically it is quite inhomogeneous. In particular
there is absolutely no observational bounds on very large variations of
primordial $^4 He$ at big distances.
It is definitely worth to obtain from astronomical data
any, even very crude limits on its mass fraction, $R_p$. Again CMB
measurements could be very helpful in this respect. Varying $R_p$ would
give rise to a different amplitudes of high multipoles at different directions
on the sky.

In the case of oscillations of the known neutrinos between themselves
primordial nucleosynthesis does not permit to put any interesting bound on
the parameters of the oscillations. However if neutrino may oscillate into
new (sterile) ones, large mixing angles and large mass differences are
excluded.

\bigskip
{\bf Acknowledgments.}
The work of A.D. was supported by Danmarks Grundforskningsfond through its
funding of the Theoretical Astrophysical Center.


\end{document}